\documentstyle[12pt,aaspp4,flushrt]{article}
\input{psfig}

\makeatletter

\@ifundefined{chapter}{\def\thebibliography#1{\section*{References\@mkboth
  {REFERENCES}{REFERENCES}}\list
  {\relax}{\setlength{\labelsep}{0em}
        \setlength{\itemindent}{-\bibhang}
        \setlength{\itemsep}{0pt}
        \setlength{\parsep}{0pt}
        \setlength{\leftmargin}{\bibhang}}
    \def\newblock{\hskip .11em plus .33em minus .07em}
    \sloppy\clubpenalty4000\widowpenalty4000
    \sfcode`\.=1000\relax}}%
{\def\thebibliography#1{\chapter*{Bibliography\@mkboth
  {BIBLIOGRAPHY}{BIBLIOGRAPHY}}\list
  {\relax}{\setlength{\labelsep}{0em}
        \setlength{\itemindent}{-\bibhang}
        \setlength{\itemsep}{0pt}
        \setlength{\parsep}{0pt}
        \setlength{\leftmargin}{\bibhang}}
    \def\newblock{\hskip .11em plus .33em minus .07em}
    \sloppy\clubpenalty4000\widowpenalty4000
    \sfcode`\.=1000\relax}}

\newlength{\bibhang}
\let\@internalcite\cite
\def\cite{\let\@citeleft(\let\@citeright)%
    \@ifstar{\citeyear}{\citefull}}
\def\citenp{\let\@citeleft\relax\let\@citeright\relax
    \@ifstar{\citeyear}{\citefull}}
\def\citefull{\def\astroncite##1##2{##1~##2}\@internalcite}
\def\citeyear{\def\astroncite##1##2{##2}\@internalcite}

\def\@citex[#1]#2{\if@filesw\immediate\write\@auxout{\string\citation{#2}}\fi
  \def\@citea{}\@cite{\@for\@citeb:=#2\do
    {\@citea\def\@citea{; }\@ifundefined
       {b@\@citeb}{{\bf ?}\@warning
       {Citation `\@citeb' on page \thepage \space undefined}}%
{\csname b@\@citeb\endcsname}}}{#1}}

\def\@cite#1#2{\@citeleft#1\if@tempswa , #2\fi\@citeright}
\def\@biblabel#1{}

\makeatother

\newcommand\approxlt{\mbox{$^{<}\hspace{-0.24cm}_{\sim}$}}
\def \angstrom{\stackrel{\rm o}{\rm A}}

\begin{document}

\title{Recovery of the Shape of the Mass Power Spectrum from the
Lyman-alpha Forest} 
\author{Lam Hui\altaffilmark{1}}
\affil{NASA/Fermilab Astrophysics Center\\ Fermi
National Accelerator Laboratory, Batavia, IL 60510}
\authoremail{lhui@fnal.gov}
\altaffiltext{1}{e-mail: \it
lhui@fnal.gov}

\begin{abstract}
We propose a method for recovering the shape of the
mass power spectrum on large scales from the
transmission fluctuations of the Lyman-alpha forest, which takes into
account directly redshift-space distortions.
The procedure, in discretized form, involves the inversion of a 
triangular matrix which projects the mass power spectrum in 3-D real-space to
the transmission power spectrum in 1-D redshift-space. We illustrate the 
method by performing a linear calculation relating the two.
A method that does not take into account redshift-space anisotropy
tends to underestimate the steepness of the mass power spectrum, in the case
of linear distortions. The issue of the effective bias-factor for the
linear distortion kernel is discussed.
\end{abstract}


\section{Introduction}
\label{intro}

In an elegant paper, Croft et al. \cite*{croft98} introduced a method
for recovering the shape of the three-dimensional primordial mass
power spectrum on 
large scales from the one-dimensional transmission power spectrum of the
Lyman-alpha forest. 
They observed that the two are related by an integral of the form:
\begin{equation}
P (k_\parallel) \propto \int_{k_\parallel}^\infty {\tilde P} {k dk \over {2
\pi}}
\label{croft}
\end{equation}
where $k_\parallel$ is the wave-vector along the line of sight,
$k$ is the magnitude of the three-dimensional wave-vector, and
$P$ and $\tilde P$ are the one-dimensional redshift-space transmission
power spectrum and the
three-dimensional redshift-space mass power spectrum respectively. 
It was suggested that redshift distortions merely change the
normalization of $\tilde P$ from its real-space counterpart, 
and so a simple differentiation of $P$ would suffice in 
recovering the shape of the three-dimensional real-space mass power spectrum.
\footnote{Croft et al. \cite*{croft98} in fact differentiated the
Gaussianized transmission power spectrum rather than the transmission
power spectrum itself. Their investigation seems to indicate that 
the two give very similar results, except that the former yields
smaller error-bars. We will consider the non-Gaussianized version of
their method in this paper for simplicity.}

Redshift distortions (see \citenp{hamilton97} and references therein),
however, imply that $\tilde P$ is in general a function of
$k_\parallel$ as well as $k$,
in which case differentiation of $P$ alone
would not recover the true shape of the three-dimensional real-space
mass power spectrum. 

We show in \S \ref{general} how to perform the inversion from the
one-dimensional redshift-space transmission power spectrum to the
three-dimensional real-space
mass power spectrum correctly, for
general, not necessarily linear, redshift
distortions. It involves the inversion of a triangular matrix, which
acts as a distortion kernel.
We illustrate the method in \S \ref{perturb} with a
perturbative example (i.e linear distortions), and demonstrate that
the method of simple differentiation generally outputs a
real-space mass power spectrum which is flatter than the true one. We
end with some concluding remarks in \S \ref{conclude}.

Before we proceed, however, let us clarify our notation on the various
power spectra treated in this paper.

\section{A Note on Notation}
\label{notation}

To avoid a proliferation of superscripts and subscripts, we adopt the
following convention for the various power spectra, $P$, discussed in this
paper. We use $\tilde{}$ to distinguish between one-dimensional and
three-dimensional power spectra: $P$ is 1-D and $\tilde P$ is 3-D
(i.e. $P$ has a dimension which is the cube-root of that of $\tilde P$).
To distinguish between the three-dimensional redshift-space
(anisotropic) versus the three-dimensional 
real-space (isotropic) power spectra, we rely on either the context or
explicit arguments of the 
power spectra: the former is denoted by $\tilde P(k_\parallel, k)$ while
the latter, being isotropic, is denoted simply by $\tilde P(k)$. 
In this paper, all one-dimensional power spectra, on the other hand,
are implicitly in redshift-space.
Finally, to tell apart the power spectrum of density from that of
transmission/flux, we use superscripts: $P^\rho$ versus $P^f$, where $\rho$
denotes the density and $f$ the transmission.

\section{General Non-perturbative Formula}
\label{general}

The three-dimensional, generally anisotropic, power
spectrum of some random field is related 
to its one-dimensional projection through the following integral
(\citenp{kp91})
\begin{equation}
P (k_{\parallel}) = \int_{k_{\parallel}}^\infty {\tilde P}
(k_{\parallel}, k) {k dk \over {2 \pi}}
\label{projection}
\end{equation}
where $k_{\parallel}$ is the wave-vector along the line of sight, and $k$
is the magnitude 
of the three-dimensional wave-vector i.e. $k^2 = k_{\parallel}^2 +
k_{\perp}^2$ where $k_{\perp}$ is the magnitude of the wave-vector
perpendicular to the line of sight.
We assume that $\tilde P$ is independent of the direction
of ${\bf k_{\perp}}$, by azimuthal symmetry, as is in the
case of redshift distortions.
Note that we have used ${\tilde P}$ for the three-dimensional power
spectrum, to distinguish it from $P$, its one-dimensional counterpart.

The power spectra are related to the three-dimensional, generally
anisotropic, two-point
correlation function $\xi$ by
the following:
\begin{eqnarray}
\label{xi}
P (k_{\parallel}) &=& 2 \int_{0}^\infty
\xi(u_{\parallel},0) \, 
{\rm cos} (k_{\parallel} u_{\parallel}) \, d u_{\parallel} \\ \nonumber
{\tilde P} (k_{\parallel},k) &=& 4 \pi \int_{0}^\infty
\int_{0}^\infty 
\xi(u_{\parallel}, u_{\perp}) \, 
{\rm cos} (k_{\parallel} u_{\parallel}) \, {J}_0 (k_{\perp} u_{\perp})
\, 
u_{\perp} d u_{\perp} d u_{\parallel}
\end{eqnarray}
where ${J}_0 (r)$ is the zeroth order Bessel function. 
The two-point correlation $\xi$ depends on the magnitude of ${\bf
u_{\perp}}$ but not its direction, again by azimuthal symmetry.

The $u_{\parallel}$ coordinate above stands for the velocity
along the line of sight (in ${\rm km \, s^{-1}}$) i.e. 
$u_{\parallel} \equiv c (\lambda-\bar \lambda) /\bar\lambda$
where $\lambda$ is the observed wavelength, $\bar\lambda$ is the mean
wavelength of interest, and $c$ is the speed of light.

The $u_{\perp}$ coordinate stands for the 
transverse distance in velocity units i.e. $u_{\perp} \equiv  \bar H
x_{\perp} / (1+\bar z)$, where $x_{\perp}$ is the actual comoving
transverse distance, $\bar z$ is the mean redshift of interest
and $\bar H$ is the Hubble parameter at that redshift.
The mean redshift and the mean wavelength
are related by $\bar\lambda = \lambda_\alpha (1+ \bar z)$,
$\lambda_\alpha = 1216 \angstrom$. 

The Fourier counterparts of $u_\parallel$ and $u_\perp$ are
$k_{\parallel}$ and $k_{\perp}$. Occasionally, we will abuse
the notation by using the ($u_{\parallel}$, $k_{\parallel}$)
pair to denote the coordinates in wavelength units i.e.
$(\lambda - \bar \lambda)$ and its Fourier transform.  

The effect of redshift-space distortions on the power spectrum, at both
small and large scales, can be described by:
\begin{equation}
{\tilde P} (k_{\parallel},k) = W(k_{\parallel}/k,k) {\tilde P} (k) \, 
\label{distortion}
\end{equation}
where ${\tilde P} (k)$ is the isotropic power spectrum in the absence of
peculiar motion, and $W$ is a suitable distortion kernel. 
Note that we rely on explicitly displaying the arguments to
distinguish between the isotropic and the anisotropic power spectra.

Finally, putting eq. (\ref{distortion}) into eq. (\ref{projection}), 
it can be seen that the one-dimensional redshift-space power spectrum
is related to the isotropic three-dimensional real-space power 
spectrum by a linear integral equation:
\begin{equation}
P (k_{\parallel}) = \int_{k_{\parallel}}^\infty
W(k_{\parallel}/k,k) {\tilde P} (k) {k dk
\over {2 \pi}} 
\label{projection2}
\end{equation}

Thus far, we have not specified the actual random field whose power
spectrum we are interested in. The random field could be
the mass overdensity $\delta = 
\delta\rho/\bar\rho$ or the transmission/flux overdensity $\delta_f =
\delta f/\bar 
f$, where $f = e^{-\tau}$, $\bar f = \langle f \rangle$,
$\delta f = f - \bar f$, and $\tau$ is the optical depth.
We will use $P^\rho$ or ${\tilde P}^\rho$ to denote
the mass power spectrum and $P^{f}$ or ${\tilde P}^f$ to denote the
transmission power spectrum.

The one-dimensional redshift-space transmission power spectrum can
also be related to 
the three-dimensional real-space mass power spectrum by an effective
kernel, which we will call $W^{f\rho}$:
\begin{equation}
P^f (k_{\parallel}) = \int_{k_{\parallel}}^\infty
W^{f\rho}(k_{\parallel}/k,k) {\tilde P}^\rho (k) {k dk
\over {2 \pi}} 
\label{projection2b}
\end{equation}

In discretized form, this is equivalent to:
\begin{equation}
{\bf P^f } = {\bf A} \cdot {\bf  {\tilde P^\rho}}
\label{projection3}
\end{equation}
where the power spectra are represented as vectors and
${\bf A}$ is an upper (or lower) triangular matrix, which is invertible
if none of the diagonal entries of ${\bf A}$ vanishes.
The special case considered by Croft et al. \cite*{croft98}
corresponds to $W^{f\rho} = \, {\rm const.}$, where inverting the
above matrix equation is
equivalent to 
the differentiation of $P^f (k_{\parallel})$. 

The problem of eq. (\ref{projection3}) is of course that $P^f
(k_{\parallel})$, for 
any given $k_{\parallel}$, depends on an infinite vector: ${\tilde P^\rho}
(k)$ for all $k$'s, from $k_{\parallel}$ to, in principle, $\infty$.
To make it useful for computation, we have to truncate the infinite
vectors somehow. Suppose one is given 
a finite vector of $P^f (k_{\parallel})$, for
$k^A_\parallel \le k_\parallel \le k^B_\parallel$ say.
Eq. (\ref{projection3}), in component form, can be rewritten as:
\begin{equation}
P^f (k_\parallel) - \Delta = \sum_{k=k_\parallel}^{k^B_\parallel}
A(k_\parallel,k) \, \tilde P^\rho (k) 
\label{projection4}
\end{equation}
where $\Delta = \int_{k^B_\parallel}^\infty W^{f\rho}(k_{\parallel}/k,k)
{\tilde P^\rho} (k) {k dk /{2 \pi}}$, and
$A(k_\parallel,k) = W^{f\rho} (k_{\parallel}/k,k) {k dk/2 \pi}$.
$A(k_\parallel,k)$ can be regarded as a triangular matrix in the sense
that $A(k_\parallel,k)$ can be set to zero for $k < k_\parallel$ by
the virtue of the lower limit of summation in eq. (\ref{projection4}).

By inverting eq. (\ref{projection4}), we can in principle determine
$\tilde P^\rho (k)$, with $\Delta$ left as a free parameter. 
We can do better, however, by the following observation:
since $\tilde P^\rho (k)$ is generally a rapidly decreasing function of $k$ for
sufficiently high $k$'s ($\sim k^{-3}$, or faster if $\rho$ is equated
with the baryon density, see footnote in \S \ref{largescales}), assuming 
$W^{f\rho} (k_{\parallel}/k,k)$ does not increase significantly with $k$, one
can see that 
$\Delta$ can be made small by choosing a sufficiently high truncation
$k^B_\parallel$. Therefore, inverting eq. (\ref{projection4}) by ignoring
$\Delta$ altogether would still give accurate estimates of
$\tilde P^\rho (k)$ for $k$'s sufficiently smaller than $k^B_\parallel$.
We will illustrate this with an explicit
example of ${\bf A}$ or $W^{f\rho}$ in the next section.

\section{A Perturbative Example}
\label{perturb}

In this section, we will perform a linear calculation of $P^f$, and we
will assume the actual shape of $P^f$ on large scales, even in the
presence of nonlinearities on small scales, agrees with that of the
linear prediction, while its amplitude might not. 
This is in the spirit of Croft et al. \cite*{croft98} who argued that,
ignoring redshift-distortions, $P^f$ should be proportional to
the linear $P^\rho$ on large scales, even though the mass fluctuations 
have gone nonlinear on small scales.
The reader is referred to Scherrer \&
Weinberg \cite*{sw97} for arguments on why that is reasonable,
in the context of local-biasing (the mapping from $\delta$ to the
optical depth or transmission can be seen as some kind of
local-biasing; see \S \ref{conclude} for subtleties however).

The output of our calculation would be a distortion kernel
$W^{f\rho}$ (eq. [\ref{projection2b}]), which may or may not be the
true kernel on large scales 
if the mass density field has already gone nonlinear on small scales.
We will have some more to say about this in \S \ref{conclude}.
Nonetheless, it is unlikely that the true $W^{f\rho}$ is equal to a
simple constant i.e. the general method of  
inverting a triangular matrix outlined in eq. (\ref{projection3}) should
be used, rather than mere differentiation. The perturbative example
set forth in this section 
should be seen as an illustration of the method. 

\subsection{Linear Fluctuations}
\label{linear}

To derive the linear theory limit of $P^f$, let us start with the
following general expression for the optical depth
(see e.g. \citenp{mer93,hui97a}):
\begin{equation}
\tau(u_{\parallel}) = \sum \int {n_{\rm HI} \over {1+\bar z}}
\left\vert {ds_{\parallel}\over 
dx_{\parallel}}\right\vert^{-1} 
\sigma_\alpha ds_{\parallel} \,\,  , \,\, \sigma_\alpha = \sigma_{\alpha 0}
{c\over {{b_T}\sqrt{\pi}}} 
\, {\rm exp} [{-{(s_{\parallel}-u_{\parallel})^2/{b_T}^2}}] \, ,
\label{tau}
\end{equation}
where $n_{\rm HI}$ is the proper number density of neutral hydrogen,
$\bar z$ is the mean redshift of interest, $x_{\parallel}$ is the
comoving spatial 
coordinate and the integration is done over the velocity
${s_{\parallel}}$ along the 
line of sight. Velocity is related to distance by
$s_\parallel = x_\parallel \bar H/ (1 + \bar z) + v_{\rm pec} (x_\parallel)$
where $v_{\rm pec}$ is the peculiar velocity along the line of sight,
and $x_\parallel = 0$ is the position where the redshift due to the
Hubble expansion alone coincides exactly with $\bar z$.
The Jacobian $|ds_{\parallel}/d{x_{\parallel}}|^{-1}$
multiplying the 
proper density 
$n_{\rm HI}$  gives us the neutral hydrogen density in velocity-space,
and the summation 
is over multiple streams of $x_{\parallel}$'s at a given ${s_{\parallel}}$.
 
The thermal profile is given in the second equality, with
$\sigma_{\alpha 0}$ being the Lyman-alpha cross section constant
(\citenp{rybicki79}). 
The width of the profile is $b_T = \sqrt{2 k_B T / m_p}$
where $T$ is the temperature of the 
gas, $k_B$ is the Boltzmann constant and $m_p$ is the mass of a
proton. 

Three pieces of physics remain to be specified if one were
to relate the optical depth and the mass distribution:
1) ionization equilibrium implies that $n_{\rm HI} \propto
[1+\delta_b]^2 T^{-0.7}$, where $\delta_b$ is the baryon overdensity;
2) the temperature-density relation $T = T_0
(1+\delta_b)^{\gamma-1}$, where $T_0$ is the mean temperature 
at $\delta_b=0$ and $\gamma$ is determined by reionization history
(\citenp{hui97b}); 3) the baryon distribution is smoothed on small
scales with respect to the mass distribution ($\delta_b
\leftrightarrow \delta$; see below). 

Without giving further details (see \citenp{hr97}), in the weak
perturbation limit ($\delta_\tau \ll 1$ where $\delta_\tau = (\tau -
\bar\tau)/\bar\tau$), it can be shown that: 
\begin{eqnarray}
\label{deltatau}
&&\delta_\tau (u_\parallel) =
\int \Biggl[
[2-0.7(\gamma-1)]\delta_b - {\partial v_{\rm 
pec}\over \partial s_\parallel} + (\gamma-1) {b_{T_0}^2 \over 4}
{\partial^2 \delta_b \over 
\partial {s_\parallel}^2}
\Biggr] W(s_\parallel - u_\parallel) ds_\parallel \, \, , \nonumber \\
&&W(s_\parallel - u_\parallel) 
\equiv {1\over   
{b_{T_0}} \sqrt{\pi}} \, {\rm exp}
[-{(s_\parallel - u_\parallel)^2/b_{T_0}^2}] \, ,
\end{eqnarray} 
where $b_{T_0} = \sqrt{2
k_B T_0 / m_p}$ is the thermal broadening width at temperature $T_0$,
and $W$ is simply a Gaussian smoothing window.

Using the fact that $\delta_f \propto \delta_\tau$ to linear order
($\delta_f = (f - \bar f)/\bar f$, with $f = e^{-\tau}$), we can
deduce the one-dimensional power spectrum of the transmission:
\begin{equation}
P^f (k_\parallel) = A \, {\rm exp} [- {k_\parallel^2/
{k^s_\parallel}^2}] \int_{k_\parallel}^{\infty} \Biggl[
[2-0.7(\gamma-1)] + f_{\Omega} {k_\parallel^2 \over k^2} -
{{\gamma-1}\over 4} k_\parallel^2 b_{T_0}^2 \Biggr]^2 {\tilde P}^\rho (k)
e^{- {k^2 / k_F^2}}
{k dk \over 2 \pi}
\label{Pf}
\end{equation}
where $f_{\Omega} = d\, {\rm ln} D/d\, {\rm ln} a$ with $D$ being the linear
growth factor and $a$ the Hubble scale factor (see \citenp{peebles80}).
The three-dimensional isotropic real-space  
mass power spectrum is denoted by ${\tilde P}^\rho (k)$, and
$k_F$ is the scale of smoothing due 
to baryon-pressure i.e. ${\tilde P}^\rho (k) \, {\rm exp} [- {k^2
/k_F^2}]$ gives 
the power spectrum of the baryons.
As argued by Gnedin \& Hui \cite*{gh98}\footnote{The $k_F$ here is
equal to the $k_F$ in Gnedin \& Hui divided by $\sqrt 2$.}, $k_F^{-1}$ should 
be given by $\sqrt 2 \bar H (1+\bar z)^{-1} f_J^{-1} x_J$, where
$x_J$ is commonly known as the Jeans scale. 
The latter is equal to ${\gamma k_B T_0 / 4 \pi a^2 G
\bar \rho \mu}$, where $\mu$ is the mean
mass per particle and $\bar \rho$ is the mean mass density. 
The numerical factor $f_J$ relating $k_F^{-1}$ and $x_J$ should
be $O(1)$, its precise value
depending somewhat on the reionization history (\citenp{gh98}), but it should
have an insignificant effect on our work here, because
we are interested primarily in the large scale fluctuations.

The other smoothing scale $k^s_\parallel$ should be equal to 
${\sqrt 2} /b_{T_0}$ due to thermal broadening, but we
can allow it to be more general to include the effect of
finite resolution as well:
\begin{equation}
k^s_\parallel = {{\sqrt 2} \over {b_{\rm eff}}} \, \, , \, \, b^2_{\rm
eff} = b^2_{T_0} + {{\rm FWHM}^2 \over {4 \, {\rm ln} 2}}
\label{ks}
\end{equation}
where ${\rm FWHM}$ is the resolution full-width-half-maximum.

The proportionality constant $A$ for eq. (\ref{Pf}) should be equal to
$\bar \tau^2$ within the context of linear theory.
However, in the spirit of Croft et al. \cite*{croft98}, 
we assumes the linear prediction gives the right shape
but not necessarily the right amplitude
for the power spectrum on large scales (see \S \ref{conclude} for
discussions). Hence, $A$ will be
left as a free constant.

Note that the integrand in eq. (\ref{Pf}) is precisely 
of the form shown in eq. (\ref{projection}). 
In fact, on large scales (small $k_\parallel$ as well as $k$ i.e.  small
compared to $k^s_\parallel$, $1/b_{T_0}$ and $k_F$), modulo
multiplicative factors, 
it reduces to the famous Kaiser \cite*{kaiser87} result, 
if one identifies $[2-0.7(\gamma-1)]$ with the usual
galaxy-bias-factor.
Interestingly, the smoothing factor ${\rm exp} [- {k_\parallel^2/
{k^s_\parallel}^2}]$ is exactly of the form commonly used
to model nonlinear redshift distortions on small scales (e.g.
\citenp{pd94,ht95}, but see also \citenp{fisher94,cfw94}). We will take advantage of this fact, and
estimate the
effect of small scale distortions on the inversion procedure
at large scales by allowing 
$k^s_\parallel$ to vary.

\subsection{Inversion on Large Scales}
\label{largescales}

Motivated by eq. (\ref{Pf}), we consider the following inversion
problem: how to estimate $\tilde P^\rho$, on large scales, from
$P^f$, for
\begin{equation}
P^f (k_{\parallel}) = \int_{k_{\parallel}}^\infty W^{f\rho} (k_{\parallel}/k,k)
{\tilde P^\rho} (k) {k dk \over {2 \pi}}
\label{inverse}
\end{equation}
where
\begin{eqnarray}
\label{Wfull}
W^{f\rho} (k_{\parallel}/k,k) =&& A' \, {\rm exp} [- {k_\parallel^2/
{k^s_\parallel}^2}] \, {\rm exp} [-{k^2 / k_F^2}] \\ \nonumber
&& \Biggl[ 1 + {f_{\Omega}
\over {2 - 0.7 (\gamma-1)}} {k_\parallel^2 \over k^2} -
{{\gamma-1}\over {4 [2 - 0.7 (\gamma-1)]}} k_\parallel^2 b_{T_0}^2 \Biggr]^2
\end{eqnarray}
where $A'$ is a constant.\footnote{Note that an alternative would be to
group the baryon-smoothing factor ${\rm exp} [-{k^2 / k_F^2}]$
together with $\tilde P^\rho (k)$ instead of 
with the rest of the terms in the distortion kernel $W^{f\rho}$.
Our inversion procedure can then be viewed as an attempt to recover
the baryon power spectrum ${\tilde P^\rho} (k) {\rm exp} [-{k^2 /
k_F^2}]$ rather than the mass power spectrum itself ${\tilde P^\rho}
(k)$. However, the two coincide on large scales.}

The above $W^{f\rho}$ is the actual distortion kernel we will
use to compute $P^f$ for some given input $\tilde P^\rho$.
However,
for the inversion problem ($P^f \rightarrow \tilde P^\rho$), we will
not assume we know 
all the parameters in $W^{f\rho}$, or even the precise
form of $W^{f\rho}$, except that, on large scales, it is equal to 
\begin{equation}
W^{f\rho}_\ell (k_{\parallel}/k,k) = A' \Biggl[ 1 + \beta_f
{k_\parallel^2 \over k^2} \Biggr]^2 \, \, , \, \, \beta_f = 
{f_{\Omega}
\over {2 - 0.7 (\gamma-1)}}
\label{Wlinear}
\end{equation}
The $\beta_f$ here is the analog of the usual $\beta$ discussed
in the context of galaxy-redshift-distortions, and $2-0.7(\gamma-1)$ is
the equivalent here of the galaxy-bias factor.
Note that the scales of our interest are much larger than
the thermal broadening width, hence the dropping of the term involving
$b^2_{T_0}$.

In other words, for the inversion problem, we assume we know the
distortion kernel in the linear regime (eq. [\ref{Wlinear}],
with the single parameter $\beta_f$), but otherwise do not have any other
information regarding 
the full distortion kernel $W^{f\rho}$ (eq. [\ref{Wfull}]) on
small scales. 
This is intended to mimic the real-life situation we find
ourselves in: that we understand linear distortions rather well, but
do not have a good grasp of nonlinear distortions on small scales.
We will use the extra parameters in the full kernel (which we ``pretend''
we do not know in the inversion procedure) to simulate the
effect of nonlinear distortions on our inversion procedure (in
particular, the parameter 
$k^s_\parallel$, which coincides with a factor commonly used
to model nonlinear distortions in galaxy surveys; see e.g. \citenp{pd94}).

Let us split the integral in eq. (\ref{inverse}) into two parts,
a part that we think we understand based on perturbation
theory, and a part that takes care of the small scale distortions
which we do not necessarily have a good handle on:
\begin{equation}
P^f (k_{\parallel}) = \int_{k_{\parallel}}^{k_\star} W^{f\rho}_\ell
(k_{\parallel}/k,k) {\tilde P^\rho} (k) {k dk \over {2 \pi}}
+ \int_{k_\star}^\infty W^{f\rho} (0,k) {\tilde P^\rho} (k) {k dk \over
{2 \pi}}
\label{split}
\end{equation}
where we have assumed: 1. $W^{f\rho} \sim W^{f\rho}_\ell$ for $k <
k_\star$; 2. $k_\parallel$ is sufficiently small so that 
$k_\parallel / k_\star \sim 0$.
The second term on the right then plays the role of $\Delta$ in eq.
(\ref{projection4}). 
As explained in the last section, the above set-up is then suitable
for an inversion analysis. 
One can imagine obtaining ${\tilde P^\rho}$ given $P^f$
for some range of $k_\parallel$'s, by inverting the matrix
$W^{f\rho}_\ell k dk/2\pi$ (which is restricted to its upper, or lower
depending on one's convention, triangular entries
by the limits of integration), treating $\Delta$ as a free parameter
or ignoring it altogether.

Instead of doing so, we will rewrite eq. (\ref{split}) into
a form that is closer to the original analysis by Croft et al.
\cite*{croft98}, thereby making manifest the differences from
our procedure suggested here.

By taking the derivative of eq. (\ref{split}) with respect to
$k_\parallel$, it can be shown that
\begin{eqnarray}
\label{iteration}
A' {\tilde P^\rho} (k=k_i) =&&  - {2 \pi \over (1+\beta_f)^2
k_i} \Biggl[ \left.{d P^f \over d
k_\parallel}\right|_{k_\parallel=k_i}  -  4 \beta_f k_i \Biggl(
\int_{k_i}^{k_{\star\star}} A' {\tilde P^\rho} (k) k^{-1} {dk \over 2 \pi} +
C_1 \Biggr) \\ \nonumber && - 4 \beta_f^2 k_i^3
\Biggl(\int_{k_i}^{k_{\star\star}} A' 
{\tilde P^\rho} (k) k^{-3} {dk \over 2 \pi} + C_2 \Biggr) \Biggr]
\end{eqnarray}
where we have used the form of $W^{f\rho}_\ell$ in eq.
(\ref{Wlinear}). The value of $k_i$ for which we will perform the
inversion would range from
some maximum $k_{\star\star}$ to whatever small $k_i$
(large scale) one might wish. The constraint is that $k_{\star\star}$ has to
be sufficiently smaller than $k_{\star}$ such that condition
number 2 as set out for eq. (\ref{split}) is satisfied.

The constants $C_1$ and $C_2$ should be
\begin{equation}
C_1 = \int_{k_{\star\star}}^{k_{\star}} A' {\tilde P^\rho} (k) k^{-1} {dk
\over 2 \pi} \, \, , \, \, C_2 = \int_{k_{\star\star}}^{k_{\star}} A' {\tilde P^\rho} (k) k^{-3} {dk
\over 2 \pi} 
\label{C}
\end{equation}

Assuming some values for $C_1$, $C_2$ and the starting wavenumber
$k_{\star\star}$, eq. (\ref{iteration}) can be used to 
obtain $A' {\tilde P^{\rho}} (k=k_i)$ for successively smaller $k_i$'s
(at, say, evenly spaced intervals).
The method adopted by Croft et al. \cite*{croft98} is equivalent
to keeping only the first term within the square brackets on the right
hand side of eq. (\ref{iteration}) i.e. a simple differentiation.

Because we are interested only in small $k_i$'s, and because ${\tilde
P^{\rho}} (k)$ generally falls rapidly with increasing $k$,
especicially at high $k$'s, we can choose $k_{\star\star}$ to be some
sufficiently large value and 
simply set $C_1 = C_2 = 0$. 
We will see that our method is robust
enough to consistently yield good agreement with the input power
spectrum on large scales (small
$k_i$), even 
though one is making an error on small scales by approximating $C_1$
and $C_2$ as zero.
Also, strictly speaking, $k_{\star\star}$ should be chosen to
be smaller than $k_\star$, which is the k-value above which the
distortion kernel is no longer described by the linear kernel in eq.
[\ref{Wlinear}] (see also conditions for eq. [\ref{split}]). We will not be
careful about it, and will see that 
one still obtains the correct $\tilde P^\rho $ on large scales, again
because $\tilde P^\rho$ falls rapidly with $k$.

For clarity let us call the method of simple
differentiation following Croft et al. \cite*{croft98} {\bf method I},
and the alternative that we propose here {\bf method II}.


Fig. \ref{pinvcomp_fid} shows a comparison of the inversion using
method I versus II. Since it is generally difficult to
judge differences between power spectral shapes in a log-log plot of
the power spectrum, we instead show the fractional error in the
inverted power spectrum as a function of $k$. The input mass power spectrum
is that of 
a SCDM (Standard Cold-Dark-Matter) universe, with $\Omega_m = 1$ and 
$h = 0.5$. 
We have chosen the parameters $\gamma = 1.5$, 
$k^s_\parallel = 0.11 ({\rm km/s})^{-1}$ and $k_F = 0.12 ({\rm
km/s})^{-1}$ in the input $W^{f\rho}$ (eq. [\ref{Wfull}]). The 
latter two values correspond to the choice $T_0 = 10^4 {\rm K}$ (eq.
[\ref{Pf}] \& 
[\ref{ks}]). (We will later alter $k^s_\parallel$ to mimic the
effect of nonlinear redshift distortions.)
The length scales shown correspond to those considered by Croft et al.
(\citenp*{croft98}). 
Since we are only interested in shapes
here, the power spectra are normalized to agree at $k = 0.005 ({\rm
km/s})^{-1}$. It can be seen that method I gives an inverted mass power 
spectrum which, on large scales, is systematically less steep, or less
steeply falling, than the 
input (i.e. the slope of the inverted power spectrum is less negative
than the actual one; see Fig. \ref{pinvp_ks} for a log-log plot of the 
power spectra).

To understand this result, let us go back to eq. (\ref{iteration}) and
rewrite it as follows, approximating $C_1$ and $C_2$ as zero
and $k_{\star\star}$ as effectively infinite:
\begin{eqnarray}
\label{checksign}
&& - {2 \pi \over (1+\beta_f)^2 k_i} \left.{d P^f \over d
k_\parallel}\right|_{k_\parallel=k_i} = A' {\tilde P^\rho} (k_i) + E
(k_i)\\ 
\label{Esign}
&& E(k_i) \equiv - {4\beta_f \over (1+\beta_f)^2}
\int_{k_i}^{\infty} A' {\tilde P^\rho} (k) k^{-1} {dk} 
- {4\beta_f^2 k_i^2 \over (1+\beta_f)^2} \int_{k_i}^{\infty} A' 
{\tilde P^\rho} (k) k^{-3} {dk} 
\end{eqnarray}
On the left hand side of eq. (\ref{checksign}) is essentially the
estimator of Croft et al. 
(\citenp*{croft98}; method I) for the shape of the mass power spectrum. 
The first term on the right is the true mass power spectrum, and
$E$ is the error of method I. Approximating a realistic
power spectrum ${\tilde P^\rho} (k)$ as $k^{-n} [1-\epsilon (k)]$
where $\epsilon(k)$ is positive and is an increasing function of $k$, 
it is not hard to show
that $d (E/{\tilde P^\rho}) / dk_i \ge 0$ by expanding to first order
in $\epsilon$, which means $E$ is
decreasing with $k_i$ slower than, or at most as fast as, ${\tilde
P^\rho}$ is. Eq. (\ref{checksign}) then tells us 
method I would systematically give a flatter estimate of the mass
power spectrum than the true one.
The limiting case of $d (E/{\tilde P^\rho}) / dk_i = 0$ occurs when
the input power spectrum ${\tilde P^\rho}$ obeys a strict power-law,
in which case $E$ has the same shape as the input, and method I
recovers the shape of the true power spectrum.

It can also be seen that both method I and II fail on small scales
(large $k$). This should come as no surprise because no attempts have been made
to model the small scale effects in the inversion procedure
laid out in eq. (\ref{iteration}) (see also eq. [\ref{split}]). We
``pretend'' that we do not 
know the actual full distortion kernel (eq. [\ref{Wfull}]), but
instead assume only knowledge of the large scale distortion
kernel (eq. [\ref{Wlinear}]) when carrying out the inversion.
Moreover,
for method II, we have not been very careful in selecting the value
of the constants $C_1$ and $C_2$: we simply set them to zero.



To estimate the effect of nonlinear distortions on our inversion
procedure, 
we decrease $k^s_{\parallel}$ to $0.028
({\rm km/s})^{-1}$, and show the outputs of method I and II
in Fig. \ref{pinvcomp_ks}. Here we are taking advantage of the fact that the
factor of ${\rm exp} [- (k_\parallel / k^s_\parallel)^2]$ in eq.
(\ref{Wfull}) is commonly used to model nonlinear distortions in the
case of galaxy surveys (see e.g. \citenp{pd94}). 
By raising the scale of nonlinear distortion by a
factor of about $4$, we hope to gain an idea of
how the as yet poorly understood nonlinear distortions on small scales might
affect the inversion of the power spectrum on large scales.
However, it remains to be checked using simulations how realistic this
choice of scale, 
or this particular parametrization of nonlinear distortions, is.
The agreement on large scales for method II is not as good as before,
but is still within about $7 \%$, and is better than that of method
I. It is possible to improve the agreement
by playing with the input parameters $C_1$, $C_2$ or $k_{\star\star}$. We
will not pursue that here.

We also show in Fig. \ref{pinvp_ks} what the inverted and input power
spectra look	 
like in a log-log plot. The subtle differences in the shapes of
the power spectra are still possible, but harder, to discern in such a plot.



Lastly, we have assumed in all tests above that the input $\beta_f$
(eq. [\ref{Wlinear}]) is known when performing the inversion.
In practice, there is an uncertainty due to the lack of knowledge of
the precise values of $\gamma$ and
$f_\Omega$. In Fig. \ref{pinvcomp_gamma}, we show a case where
the inversion $\gamma$ is chosen to be slightly different from the known input
$\gamma$. (The actual value for $\gamma$ in the real
universe is likely to have a narrow range $1.3 \, \approxlt \, \gamma
\, \approxlt \, 1.6$ c.f. 
\citenp{hui97b}). The 
impact on the recovery of the power spectrum shape appears
minimal, for the small change in $\gamma$.
Similary, $f_\Omega$ at a redshift of around $3$ should fall in a
narrow range (close to $1$), for reasonable values of $\Omega_m$ and
the cosmological constant today.


\section{Conclusion}
\label{conclude}

The main aim of this paper is to draw attention to the fact that
redshift distortions in general make the three-dimensional
redshift-space mass power
spectrum anisotropic, and so the inversion from
the projected one-dimensional redshift-space transmission power spectrum to the
three-dimensional 
real-space mass power spectrum involves more than
a simple differentiation (eq. [\ref{projection}]).
Given a kernel ($W^{f\rho}$ in eq. [\ref{projection2b}]) that relates
the relevant power spectra, it
is possible to perform the inversion by essentially inverting
a triangular matrix proportional to $W^{f\rho}$ (eq.
[\ref{projection3}] \& [\ref{projection4}]).

We have demonstrated this idea with a kernel $W^{f\rho}$ that is motivated
by linear perturbation theory. In general, we find that
a simple differentiation method tends to make the inverted three-dimensional
real-space power spectrum flatter than it really is on large scales. A
procedure that  
remedies this is outlined in eq. (\ref{iteration}). 
We have referred to the former (straightforward differentiation) as
method I, and the latter as method II. 

\begin{figure}[htb]
\centerline{\psfig{figure=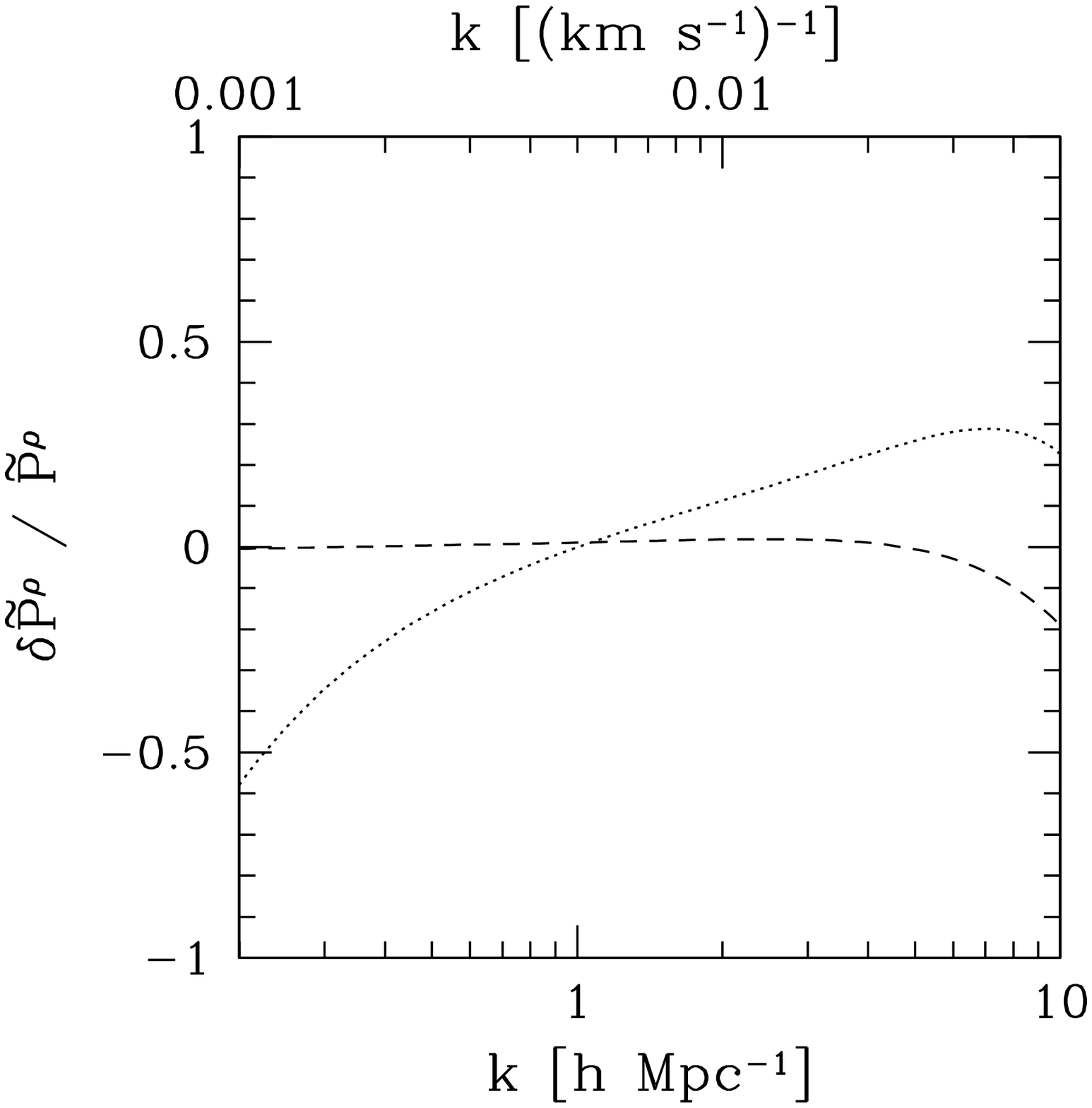,height=3.0in}}
\caption{A comparison of the Croft et al. (1998) inversion (method I; {\it
dotted line})
versus inversion set forth in eq. (\protect{\ref{iteration}}) (method
II; {\it dashed line}). On the y-axis is $\delta {\tilde P^\rho} /
{\tilde P^\rho}$ where 
$\delta {\tilde P^\rho} = {\tilde P^\rho}_{\rm output} - {\tilde
P^\rho}$, and ${\tilde P^\rho}$ is the input real-space mass power
spectrum, and ${\tilde P^\rho}_{\rm output}$ is the output.
The input power spectrum
is that of the SCDM model, and the input
parameters are $\gamma = 1.5$, $k^s_\parallel = 0.11 ({\rm
km/s})^{-1}$ and $k_F = 0.12 ({\rm 
km/s})^{-1}$ (eq. [\protect{\ref{Wfull}}]). The inversion parameters
are $k_{\star\star} = 0.14 ({\rm km/s})^{-1}$, $C_1 = C_2 = 0$ (eq.
[\ref{iteration}]). The inversion
$\beta_f$ is 
assumed to be the same as that in the input (i.e. $1/1.65$, see eq.
[\ref{Wlinear}]). 
The outputs are normalized to match the input at $k = 0.005 ({\rm
km/s})^{-1}$.
}
\label{pinvcomp_fid}
\end{figure}

\begin{figure}[htb]
\centerline{\psfig{figure=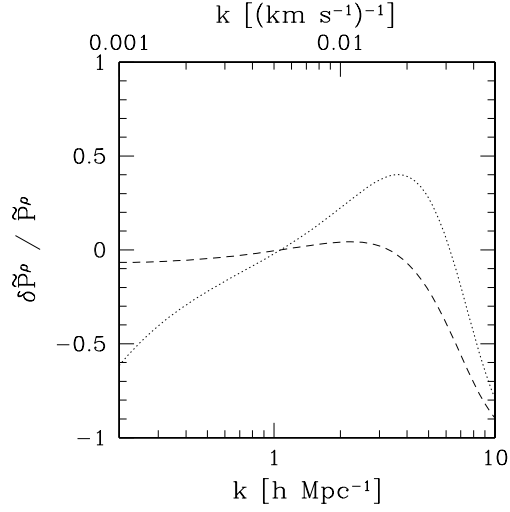,height=3.0in}}
\caption{Same as Fig. \ref{pinvcomp_fid} except that $k^s_{\parallel}
= 0.028 ({\rm km/s})^{-1}$. } 
\label{pinvcomp_ks}
\end{figure}

\begin{figure}[htb]
\centerline{\psfig{figure=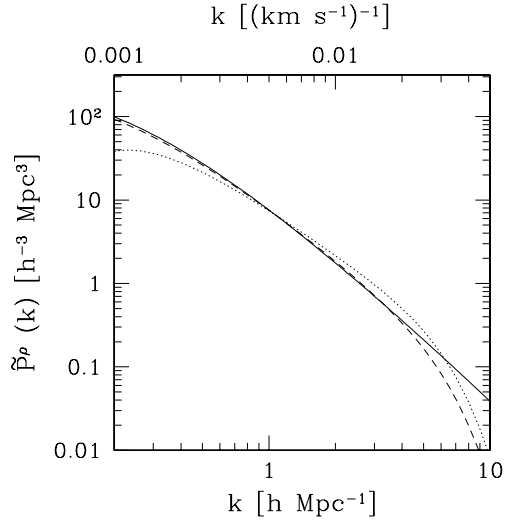,height=3.0in}}
\caption{The solid line is the input linear power spectrum.
The dashed line is the output by using method II (eq.
[\ref{iteration}]), and the dotted line is the output by using method
I i.e. simple differentiation. The parameters are the same as in Fig. \ref{pinvcomp_ks}.}
\label{pinvp_ks}
\end{figure}

\begin{figure}[htb]
\centerline{\psfig{figure=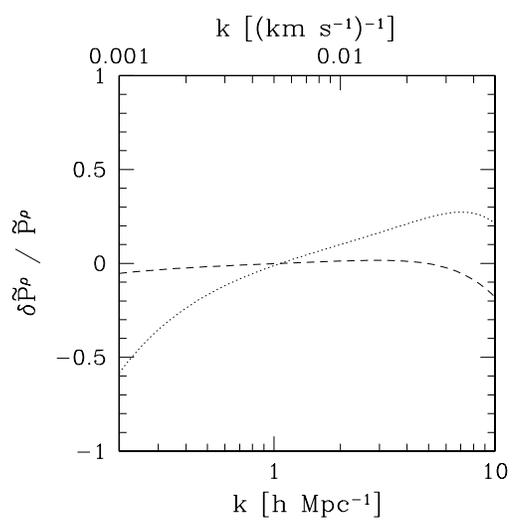,height=3.0in}}
\caption{Same as Fig. \ref{pinvcomp_fid} except that the inversion
$\gamma-1$ is chosen to be $0.3$ instead of the input value $0.5$.}
\label{pinvcomp_gamma}
\end{figure}

As we have remarked before in \S \ref{largescales}, 
if the three-dimensional real-space mass power spectrum ${\tilde P^\rho}
(k)$ obeys a strict power-law,
the simple differentiation procedure of method I will recover the same
power-law. That we observe a deviation of the method-I-inverted power
spectrum from the input in Fig. \ref{pinvcomp_fid} to \ref{pinvcomp_gamma}
is a reflection of the fact that
a realistic power spectrum (such as CDM) is only well-approximated
by a power-law for narrow ranges of $k$. The differences between
method I and II are not easy to discern in a log-log plot such as the
one shown in Fig. \ref{pinvp_ks}, especially when one is dealing
with observed or simulated data with noise. The systematic error of method I is
nonethless present, and more easily seen in plots such as
the one in Fig. \ref{pinvcomp_ks}. A cursory inspection of some of the log-log
plots of the inverted power spectrum in 
Croft et al. 
(\citenp*{croft98}) seems to indicate that method I does give a
slightly flatter power spectrum compared to the input on large scales,
but it certainly should be more carefully quantified with simulations.

An interesting consequence of redshift-space distortions is that 
the one-dimensional transmission power spectrum $P^f$ is no longer
guaranteed to be monotonically decreasing (\citenp{kp91}), unlike in
the case where 
the distortion kernel is trivial (i.e. set $W^{f\rho}$ to constant in
eq. [\ref{inverse}]). Method I, where $P^f$ is
simply differentiated to obtain the mass power spectrum $\tilde
P^\rho$, could then give a negative mass power
spectrum and fail dramatically (see eq. [\ref{checksign}], and
discussions that follow).  

For a power-law mass spectrum with $\tilde P^\rho (k) \propto k^{-n}$,
it is simple to show using the linear distortion kernel (eq.
[\ref{Wlinear}]), together with eq. (\ref{inverse}), that 
$d P^f / d k_\parallel > 0$ if
$[-\sqrt{\beta_f^2 + 6 \beta_f + 1} - (1-\beta_f)]/(1+\beta_f) <  n <
[\sqrt{\beta_f^2 + 6 \beta_f + 1} - (1-\beta_f)]/(1+\beta_f)$.
For the parameters $\Omega_m = 1$ and $\gamma = 1.6$, which were 
adopted by Croft et al. \cite*{croft98}, this implies a range of 
$ -1.62 < n < 1.17$, within which method I should give a negative
inverted mass power spectrum. Of course, a strict power-law power
spectrum with $n$ in such a
range is not very interesting because it gives a diverging
one-dimensional transmission power spectrum (
$n > 2$ is required for convergence).
It is for this reason that a direct comparison cannot be made between the above
simple estimates and a test-case shown by 
Croft et al. \cite*{croft98} in which the initial power spectrum has $n =
1$. In fact, for a scale-free initial power
spectrum of $k^{-n}$, the stable-clustering mass power 
spectrum at high $k$ should be $k^{-6/(5-n)}$ (\citenp{peebles80}).
For $n=1$, this means the high $k$ mass power spectrum asymptotes to
$k^{-1.5}$, which is 
still not enough to regularize the integral for the transmission power
spectrum (eq. [\ref{inverse}] together with eq. [\ref{Wlinear}]). It
is likely that in their test-case, 
an effective ultraviolet cut-off is imposed by the finite resolution
of the simulation, or else the true nonlinear redshift-distortion kernel
provides an effective regularization. Note, however, that there is no
small-scale Jeans smoothing to help here because Croft et al.
\cite*{croft98} used N-body-only simulations. 

Nonetheless, we should emphasize that while, on large scales, the
direction of the 
systematic error of method I is on firm ground
(i.e. systematic flattening), the size of the error, on the other
hand, is subject to
further investigation. We have used linear theory in \S
\ref{perturb} to estimate the magnitude of this effect.
The linear calculation gives us the effective ``bias-factor'' (let us call it
$b_{\beta}$, which equals $2-0.7 (\gamma-1)$ in eq. [\ref{Wlinear}])
associated with the 
distortion parameter $\beta_f$ (i.e. $\beta_f = f_\Omega /
b_{\beta}$), but this might not be the correct $b_{\beta}$ in 
the presence of nonlinear fluctuations on small scales.

For instance, according to arguments by Scherrer \& Weinberg
(\citenp*{sw97}), the galaxy-bias $b_g$ (different from $b_{\beta}$ above, see
below), defined as the ratio of the galaxy
to mass two-point function on large scales (where galaxy is related
to mass by a local transformation), generally involves all the higher
derivatives of the local transformation around $\delta = 0$ i.e. not
just the first derivative, as is in the case of a linear perturbative
calculation. 

One might naively think that, in the case of the Lyman-alpha forest,
$b_{\beta}$ should be 
set equal to the analog of the galaxy-bias defined above, namely
the ratio of the transmission to mass power spectrum
on large scales. Let us denote this latter quantity also by $b_g$.
This number is smaller than $1$ in our case (because 
the exponential ($e^{-\tau}$) suppresses large fluctuations), which means
the redshift distortions are more pronounced than we have assumed in
\S \ref{perturb}, and 
the systematic error of method I should be even larger! 

However, there are at least two reasons to suspect that this is not
the correct conclusion. First, as emphasized recently by Dekel \& Lahav
(\citenp*{dl98}), the bias factor that shows up in the linear
distortion kernel ($b_{\beta}$) is not necessarily the same as $b_g$
defined above, because of the nonlinearity of the biasing
transformation (see below for a discussion of the transformation
relevant in our case; we do not suffer from stochastic
biasing, however). In fact, according to Dekel \& Lahav, even the form of the
linear distortion kernel 
could be slightly modified (see also \citenp{pen97}). 

A second, perhaps more important, reason is 
that the mapping from the mass density field $\rho$ to the
transmission $f$ actually involves two local ``biasing'' transformations with
the redshift-space distortion in between.
First\footnote{We ignore the spatial dependence of the thermal profile
to simplify the discussion here; see eq. (\ref{tau}).}, the mass
density in real-space is related to the neutral 
hydrogen density in real-space through the local transformation $n_{\rm HI}
\propto \rho^{2-0.7(\gamma-1)}$; second, this is then ``distorted''  into
the optical depth in redshift-space through $\tau \propto
[\rho^{2-0.7(\gamma-1)}]_z$, where $[\,]_z$ denotes a quantity in
redshift-space; finally, another local transformation 
maps the optical depth in redshift-space to the transmission in
redshift-space: $f = e^{-\tau}$. It is plausible that the
first two steps determine the correct value of $b_{\beta}$, while
the last merely shifts the overall normalization of the
final redshift-space power spectrum of the transmission on large
scales (see \citenp{hui98}).

Lastly, there is of course the possibility that even the form of the
large-scale 
distortion kernel in eq. (\ref{Wlinear}) might be incorrect, or as is
more likely the case, that the linear distortion kernel applies only
at (a small range of) the largest scales (see e.g. \citenp{cfw94}). 
This is related to the question of how nonlinear clustering on small
scales, or translinear clustering on intermediate scales, affects
the large scale behaviour of redshift distortions. Couple this with
the effects of nonlinear local transformations: we indeed have a
complicated problem here. Some of these issues are begining to be
addressed (see \citenp{fn96,th96,scoc98,hks98} for the former, and
\citenp{sw97,dl98,hui98} for the latter).
The good news, at least in the case of the Lyman-alpha forest, is that
the form of the relevant local ``biasing'' transformation is known exactly.

Both issues, the problem of translinear or nonlinear redshift-space
distortions and 
the problem of biasing in real as well as in redshift-space, are
obviously of great 
interest in the wider context of large scale structure and galaxy surveys.
Analytical calculations, with reality checks using simulations,
would be necessary to address these questions. 
We hope to pursue aspects of these issues in a future publication.

As this paper was nearing completion, the author became aware of a
preprint by Nusser \& Haehnelt (\citenp*{nh98}) who also considered the
effects of redshift-space distortions on the forest, not in the case
of inversion from the transmission power spectrum to the mass power
spectrum, but in the case 
of the recovery of the mass density field itself. 
The author thanks Rupert Croft and Albert Stebbins for useful
discussions. This work was supported by the DOE and the NASA
grant NAG 5-7092 at 
Fermilab.


\newpage


\end{document}